\documentclass{elsart}
\usepackage{graphics}
\usepackage{psfig}
\usepackage{natbib}
\begin{document}
\begin{frontmatter}

\title{Possibility of Measuring the Width of Narrow Fe II Astrophysical Laser Lines in 
the Vicinity of $\eta$ Carinae by means of Brown-Twiss-Townes Heterodyne 
Correlation Interferometry }

\author[Lund]{S. Johansson}, 
\author[Troitsk]{V.S. Letokhov$^a$}

\address[Lund]{Lund Observatory, Lund University, 
            P.O. Box 43, S-22100 Lund, Sweden} 
\address[Troitsk]{Institute of Spectroscopy, Russian Academy of Sciences, Troitsk, 
     Moscow region, 142190, Russia}

%            email: Sveneric.Johansson@astro.lu.se,
%                    Vladilen.Letokhov@astro.lu.se}

\begin{abstract}We consider the possibility of measuring the true width of the narrow
Fe II optical lines observed in spectra of the Weigelt blobs in the vicinity of Eta Carinae.
The lines originate as a result of stimulated amplification of 
spontaneous emission of radiation in quantum transitions between energy levels 
showing inverted population (Johansson \& Letokhov, 2002, 2003, 2004). 
The lines should have a subDoppler spectral width of 30-100 MHz, depending on the 
geometry of the lasing volume. To make measurements with a spectral resolution of
$R>10^7$ and an angular resolution better than 0.1 arcsec, we suggest the use of
the Brown-Twiss-Townes optical heterodyne intensity correlation interferometry. 
The estimates made of the S/N ratio for the optical heterodyne astrophysical laser 
experiment imply that it is feasible.

\end{abstract}
\end{frontmatter}

\section{Introduction}

The conclusion about the existence of an Fe II astrophysical laser (APL) active 
in spectral lines in the range 0.9-–2 $\mu$m in the Weigelt blobs in the vicinity 
of $\eta$ Carinae (Johansson \& Letokhov, 2002, 2003, 2004) was drawn on the 
basis of data obtained with the $HST/STIS$ facility (Gull et al., 2001). 
The high spatial resolution of HST/STIS allows observation of the emission 
spectrum of the Weigelt blobs separated from the photospheric spectrum of the 
central source. The inference about the laser origin of some of the Fe II 
spectral lines was based on a detailed analysis of the photoselective 
excitation of the Fe$^+$ ions and the radiative decay pathways from some 
specific energy levels. These 
levels are resonantly excited by the intense H Ly$\alpha$ radiation coming from 
the HII region around the Weigelt blobs. The analysis was initiated
 by our finding of unexpectedly strong lines in the $HST/STIS$ 
spectra that indicated a transformation of a series of weak transitions into 
intense (allowed-like) lines due to an inverted population. This can occur as a 
result of stimulated emission of radiation in transitions between energy levels 
having an inverted population.

However, the maximum spectral resolution of the $HST/STIS$ instrument is  
$R\simeq 10^5$, which is an order of magnitude less than the Doppler width of the 
spontaneous Fe II lines from the cold HI region of the Weigelt blobs. As 
discussed in Sec. 2 laser spectral lines can be much narrower. Hence, it seems 
very desirable to measure the true width of these laser lines at least in the 
range 0.9--1.0 $\mu$m with the aid of ground-based telescopes. This would be a 
direct proof of the laser effect in the Fe II lines, as was the case with the 
spectral measurements in the microwave region in radio astronomy that led to the 
discovery of astrophysical masers (Elitzur, 1992).

The Brown-Twiss correlation interferometry (Hanbury Brown \& Twiss, 1956; 
Hanbury Brown, 1974), modified by heterodyne detection using a CO$_2$ laser as a 
local oscillator (Johnson et al., 1974) for the 10 $\mu$m region, is well 
suited to achieve very high angular and spectral resolution simultaneously. We 
will call this method the Brown-Twiss-Townes (BTT) technique. In the range 
0.9–-1.0 $\mu$m this can be done today by means of two spatially separated 
telescopes equipped with two heterodyne photoreceivers, e.g. avalanche
diodes, and a tunable semiconductor laser diode transporting its radiation 
via an optical fiber (Sect. 3). A general approach of this method has already 
been considered and discussed in a non-astronomical journal (Lavrinovich \& 
Letokhov, 1976; Letokhov, 1996). In the present paper, we focus on the use of 
the BBT correlation heterodyne interferometry to study the Fe II laser lines in 
the range 0.9–-1.0 $\mu$m and the appropriate signal-to-noise (S/N) ratio 
estimates (Sect. 4).

\begin{figure}
\begin{center}
    \rotatebox{270}{\resizebox{8cm}{!}{\includegraphics{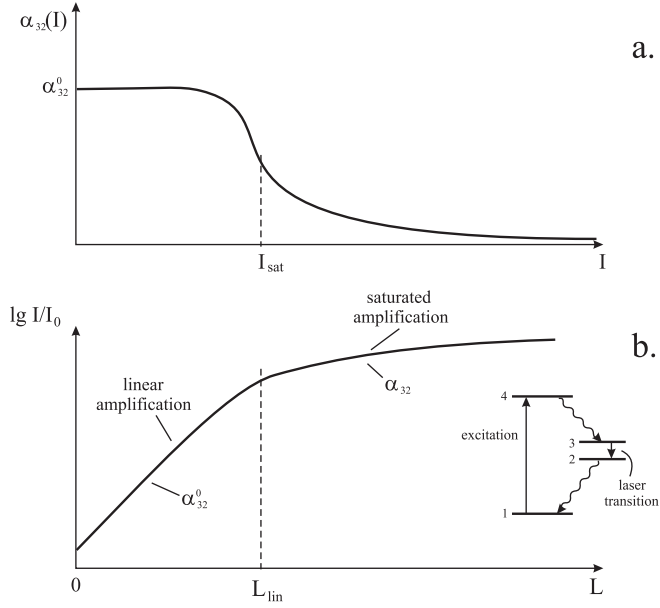}}}
\end{center}
\hfill  
     \caption{(a) Change in the amplification coefficient $\alpha_{32}$ during 
        the transition from the linear amplification regime to the saturated 
         amplification regime; 
	(b) Growth in the intensity of the amplified spontaneous radiation in 
	the linear and saturated regimes ($L_{\mathrm {lin}}$ is the length, over 
        which the linear (non-saturated) amplification and exponential growth of 
        intensity occur.
        The insert shows a schematic of the four-level radiative cycle involving
        the laser transition.}
    \label{Fig. 1}
\end{figure}

\section{Expected Spectral Width of the APL}

The astrophysical laser is a laser amplifier that intensifies weak spectral 
lines of spontaneous emission of radiation in quantum transitions between pairs 
of energy levels having an inverted populations. The linear amplification factor 
$K={\mathrm {exp}}(\alpha^0_{32}D)$, where $\alpha^0_{32}$ (cm$^{-1}$) is the 
linear gain per unit length of the laser operating in the $3\rightarrow 2$ 
transition (see Fig. 1) between inverted energy levels, and $D$ is the size of 
the amplifying region, for example, the diameter of the Weigelt blobs $D\simeq 
10^{15}$ cm (Johansson \& Letokhov, 2002, 2003, 2004). Under linear 
amplification conditions, the intensity of spontaneous radiation increases 
exponentially with the amplification length until the amplification is saturated 
(Fig. 1). Under saturated amplification conditions, the stimulated transition 
rate $W_{32}$ exceeds the pumping rate $W_{\mathrm {exc}}$ of the excited level 
{\it 3} and the gain drops as (Siegman, 1966)
%
% Eq 1
\begin{equation}
\alpha_{32}=\frac{\alpha_{32}^0}{1+(I/I_{\mathrm {sat}})},
\end{equation}
where {\it I} is the intensity of the radiation being amplified (in 
photons/cm$^2\cdot$s). $I_{\mathrm {sat}}$ is the amplification saturation 
intensity given by the expression
%
%Equation 2
\begin{equation}
I_{\mathrm {sat}}=\frac{\hbar\omega_{32}}{\sigma_{32}\tau_3},
\end{equation}
where $\tau_3$ is the lifetime of upper level $\it 3$ and $\sigma_{32}$ is the 
stimulated emission cross-section (in cm$^2$),
%
%Equation 3
\begin{equation}
\sigma_{32}=\frac {\lambda_{32}^2}{2\pi}\frac{A_{32}}{2\pi\Delta\nu_{\mathrm D}}.
\end{equation}
The Einstein coefficient for the $3\rightarrow 2$ transition is $A_{32}$, and 
$\Delta\nu_{\mathrm D}$ is the Doppler width (in Hz) of the $\lambda_{32}$ line. 
When $I>>I_{\mathrm {sat}}$, amplification varies linearly, not exponentially, 
with the amplification length (Siegman, 1986; Elitzur, 1992).

Under linear conditions, amplification takes place predominantly at the center 
of the bell-shaped Doppler profile, and the spectrum of the radiation being 
amplified narrows as (Casperson \& Yariv, 1972)
%
% Eq 4
\begin{equation}
\Delta\nu=\frac{\Delta\nu_{\mathrm D}}{\sqrt{1 + \alpha^0_{32}L_{\mathrm {lin}}}},
\end{equation}
where $L_{\mathrm {lin}}$ is the linear amplification length, i.e., the length 
of the amplifying region where no amplification saturation occurs. The further 
spectral evolution of the amplified radiation depends on its 
angular divergence, i.e. essentially on the geometry of the amplifying region.

Before we consider how various geometries of the amplification volume 
influence the spectral width of an APL it is useful to discuss how 
the divergence of the monochromatic light field affects the resonant 
interaction width $\Delta\nu_{\mathrm {int}}$ of the the spectral line, non-homogeneously 
(Doppler) broadened by the moving atoms (ions), as  
illustrated in Fig. 2 (Letokhov \& Chebotayev, 1977). 

\begin{figure}
\begin{center}
    \rotatebox{0}{\resizebox{10cm}{!}{\includegraphics{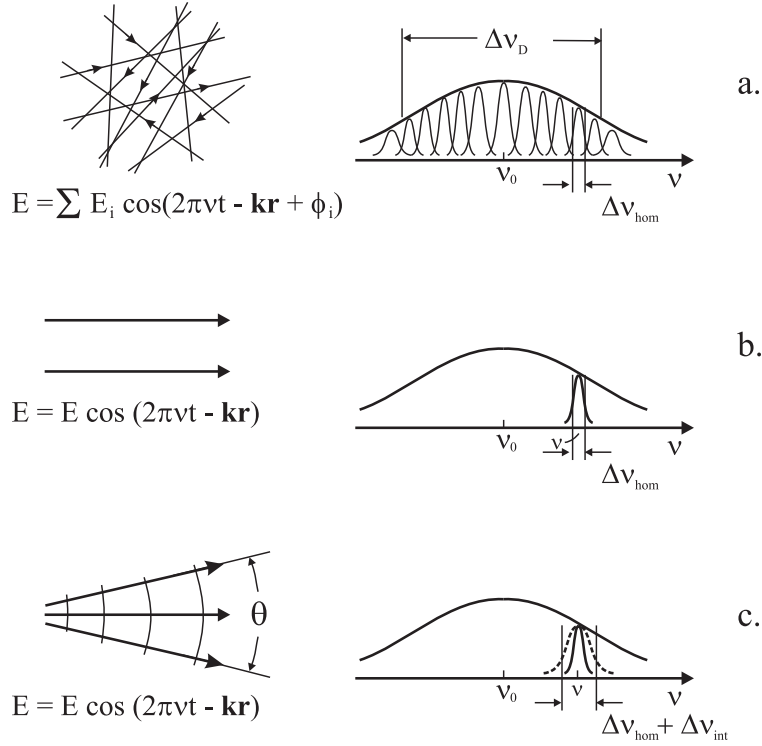}}}
\end{center}
\hfill  
     \caption{Illustration of the effect of the divergence of the monochromatic 
light field on the resonant interaction width from a non-homogeneously (Doppler) 
broadened spectral line : (a) Isotropic monochromatic field; (b) plane light wave; (c) light field with the 
angular divergence $\Theta$.}
    \label{Fig. 2}
\end{figure}

When spectral-line broadening 
is non-homogeneous, the light wave interacts only with particles with which it is in 
resonance. The portion of the particles interacting with the field depends both on the 
homogeneous spectral width $\nu_{\mathrm {hom}}$, and on the divergence of the light field. 
If a monochromatic field is isotropic, all of the particles can interact with the field, 
whatever their velocity (Fig. 2a). On the other hand, a plane travelling wave 
$E{\mathrm {cos}}(2\pi\nu t - {\bf kr})$ interacts only with particles located within the 
spectral range of the homogeneous width $\Delta \nu_{\mathrm {hom}}$ at 
the resonance frequency $\nu = \nu_0 + ({\bf {kv}}/2\pi)$ (Fig. 2b). In other words, 
the field interacts only with particles having a definite velocity projection on the 
travelling-light-wave direction: 
$\mid \nu -\nu_0 + {\bf {kv}}/{2\pi}\mid \leq \Delta \nu_{\mathrm {hom}}/2$.
In case of finite angular divergence $\Theta$ the interaction spectral width is 
proportional to $\Theta$ (Fig. 2c).

Let us for simplicity consider the following three limiting geometrical cases: 
(a) elongated volume, (b) spherical volume and (c) disk-like volume.

\begin{figure}
\begin{center}
    \rotatebox{270}{\resizebox{8cm}{!}{\includegraphics{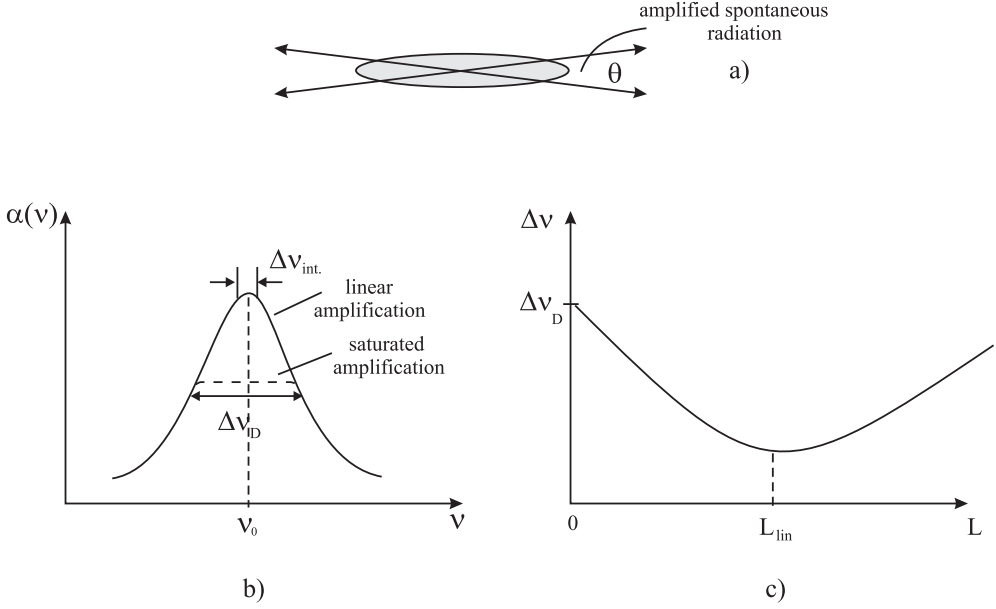}}}
\end{center}
\hfill  
     \caption{(a) Elongated shape of the amplification volume, (b) change of the 
spectral profile of the non-homogeneously broadened amplification line under 
saturated conditions, and (c) evolution of the line width $\Delta\nu$ of the spontaneous 
radiation being amplified in such a way that the spectral line changes from an initial 
narrowing to a rebroadening at $L=L_{\mathrm {lin}}$.}
    \label{Fig. 3}
\end{figure}

(a) {\em Elongated Amplification Volume} (Fig. 3a). In this case, the angular 
divergence $\Theta$ of the amplified radiation is governed by the angular spread 
of the amplifying region: 
%
% Eq 5
\begin{equation}
\Theta \simeq \frac{a}{L},
\end{equation}
where $a$ and $L$ are the lateral and longitudinal sizes of the volume, 
respectively. In rarefied gas condensations, the collisional 
broadening is very small. A more or less directed light beam interacts with the 
narrow spectral range $\Delta\nu_{\mathrm {int}}$ of the non-homogeneously broadened 
Doppler profile of the spectral line (Letokhov \& Chebotayev, 1977):
%
% Eq 6
\begin{equation}
\Delta\nu_{\mathrm {int}}= (\Theta/\pi)\cdot\Delta \nu_{\mathrm {D}},
\end{equation}

If $\Theta <<\pi$, amplification gets saturated first at the center of the Doppler 
profile, and then the amplification saturation gradually spreads over the entire profile. 
As a result, the spectral profile of saturated amplification becomes flat (Fig. 3b), 
which gives rise to the reverse effect - a spectral rebroadening of the radiation 
being amplified under saturated conditions (the spectral re-broadening effect, Fig. 
3c). This effect was analyzed by Litvak (1970) in the case of microwave space masers.

\begin{figure}
\begin{center}
    \rotatebox{270}{\resizebox{8cm}{!}{\includegraphics{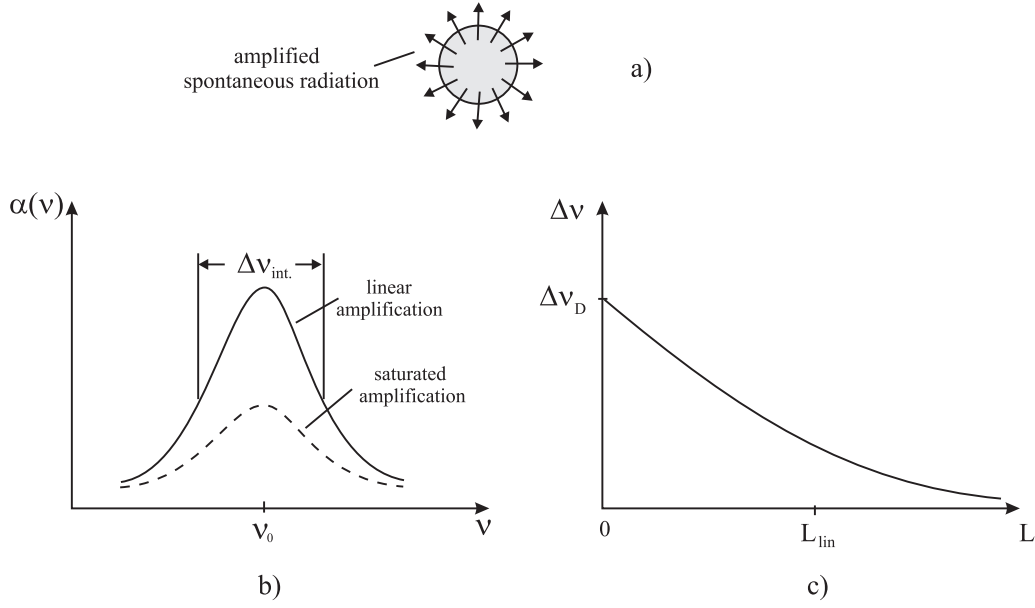}}}
\end{center}
\hfill  
     \caption{a) Spherical shape of the amplification volume, (b) absence of changes 
in the shape of the non-homogeneously broadened spectral line upon saturation by 
isotropic radiation, and (c) corresponding absence of the spectral re-broadening 
effect.}
    \label{Fig. 4}
\end{figure}

(b) {\em Spherical Amplification Volume} (Fig. 4a). With this 
geometry of the amplifying region, the amplified radiation is isotropic. The 
saturation of the Doppler profile under the effect of isotropic radiation occurs 
equally for the entire profile, i.e., without any change in the shape of the 
amplified line, as shown in Fig. 4b (Letokhov \& Chebotayev, 1977). In this 
case, the effect of rebroadening of the amplified radiation is absent 
(Fig. 4c), and the spectral width of the radiation is defined by 
%
% Eq 7
\begin{equation}
\Delta\nu= \frac{\Delta\nu_{\mathrm D}}{\sqrt{1+\alpha_{32}^0 D}},
\end{equation}
where $D$ is the full size of the amplifying region, for instance, the 
diameter of a Weigelt blob, which can be much greater than the linear 
amplification length $L_{\mathrm {lin}}$. Since the factor $\alpha_{32}^0 D$
 can be as great as 100, the width of the amplified line can be an order of 
 magnitude smaller than $\Delta \nu_{\mathrm D}$. 
 
(c) {\em Disk-Like Amplification Volume.} This is an intermediate case between 
(a) and (b). In the direction along the disk plane, amplification saturation takes 
place under the effect of radiation with a geometrical angular divergence of 
$\Theta \sim a/b$, where $a$ is the thickness of the amplifying disk and $b$ is 
its diameter. When $\Theta<<\pi$ amplification should be saturated first in the 
narrow central region of the Doppler profile after which the top of the profile 
should become flat, as in the case with an elongated volume (a). Thus, in 
the present case (c) the width of the laser line should be comparable with the 
Doppler width $\Delta\nu_{\mathrm D}$. However, if saturation is attained in the 
direction perpendicular to the disk plane as well, the "quasi-isotropic" radiation 
will provide for saturation of the entire Doppler profile, as in the case 
with a spherical amplifying volume (b). One could then expect that the laser 
line rebroadening effect will be absent, so that the line will narrow in 
accordance with Eq. (7):
%
% Eq 8
\begin{equation}
\Delta\nu \simeq \frac{\Delta\nu_{\mathrm D}}{\sqrt{1+\alpha_{32}^0 a}}.
\end{equation}

For astrophysical lasers active in the wavelength range 0.9--1.0 $\mu$m in the 
Weigelt blobs of $\eta$ Carinae, any of the different types of geometry of the 
amplifying region mentioned above can be realized, either (a), or (b), or 
else (c). The various types of geometry that can exist for different sets 
of parameters characterizing the Weigelt blob will be considered in a future 
publication. What is important at the moment is that the laser lines observable 
in the range $0.9–-1.0 \mu$m can have a width, $\Delta\nu$, somewhere between 
$\Delta\nu_{\mathrm D}$ and down to at least $0.1\cdot\Delta\nu_{\mathrm D}$. 
The magnitude of $\Delta\nu_{\mathrm D}$ for the Fe II lines from the HI region 
of the blobs depends on the temperature $T$ and can amount to 
$\Delta\nu_{\mathrm D}\sim 300- 1000$ MHz, since $T$ is in the range 100 to 1000 
K. Accordingly, the laser line width $\Delta\nu$ can lie between 30 and 1000 
MHz. To measure such lines adequately requires a spectral resolution of $R\sim 
10^7$, which is very difficult to achieve with the standard spectroscopic 
techniques. For this reason, it is expedient to use the Brown-
Twiss-Townes correlation method, modified by utilizing the up-to- date capabilities 
of electronics, optics, and lasers.

\section{Brown-Twiss-Townes Optical Heterodyne Intensity Correlation Interferometry 
with Very High Spatial and Spectral Resolution}

In the 1950-1960's, R. Hanbury Brown and Q. Twiss proposed and developed a new 
method under the name {\em intensity interferometry} (Hanbury Brown, 1974), which is 
characterized by its very high spatial (angular) resolution. It produced accurate 
diameter measurements of 36 bright stars, using a pair of 65-meter collectors 
and a 188-meter baseline. The method was perhaps the first among a series of 
new techniques based on advanced technology that led to a substantial progress in 
attaining levels of high and ultrahigh resolution in astronomy (see review by Saha, 
2002). For example, the speckle interferometry technique helped to discover gas 
condensations (blobs) in the vicinity of Eta Carinae (Weigelt \& Ebersberger, 
1986). To master new wavelength regions and achieve high spectral resolution 
levels, the intensity interferometry method was modified to become {\em heterodyne 
interferometry} (Johnson et al., 1974). This technique uses a local monochromatic 
laser oscillator to produce beats between the light wave of the star of interest 
and the coherent laser wave of the local oscillator. The method can be 
considered intermediate between intensity interferometry and direct 
interferometry. {\em Townes} and co-workers made successful observations in the 10-$\mu$m 
infrared window of the atmosphere using a CO$_2$ laser as a local oscillator 
(Johnsson et al., 1974; Townes, 1977). The baseline used consisted of a pair of 
auxiliary telescopes spaced a few meters apart at Kitt Peak's solar telescope.

\begin{figure}
\begin{center}
    \rotatebox{270}{\resizebox{5cm}{!}{\includegraphics{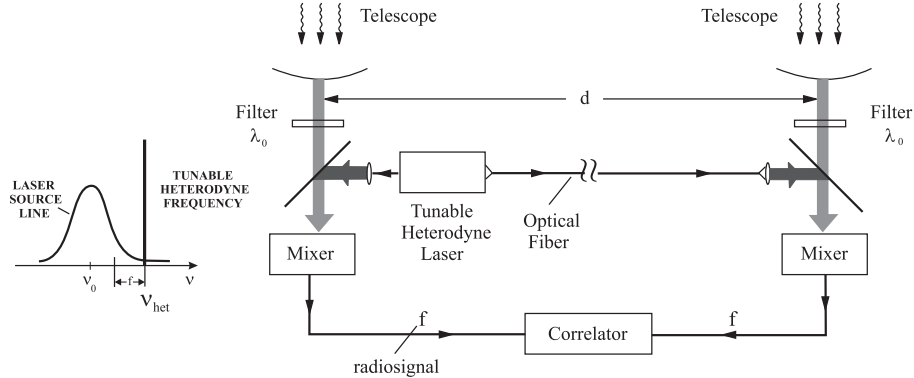}}}
\end{center}
\hfill  
     \caption{Brown-Twiss-Townes optical laser heterodyne intensity correlation 
interferometer with two separated telescopes, local tunable laser diode 
heterodyne of frequency $\nu_{\mathrm {het}}$.}
    \label{Fig. 5}
\end{figure}

In Fig. 5 we present a schematic diagram of a Brown-Twiss-Townes (BTT) optical 
heterodyne interferometer that can be used to measure both the angular size and 
emission spectrum of Fe II laser line sources in the Weigelt blobs of $\eta$ 
Carinae. A specific feature of such a correlation interferometer is the use of a 
0.9-1 $\mu$m tunable monomode diode laser as a local oscillator and an optical 
fiber to transport this monomode laser radiation. This is much easier to 
accomplish than to transport the space laser radiation received by the 
telescopes. The distance $d$ between the telescopes should meet the 
requirement in angular resolution that the the radiation emitted by the 
Weigelt blob studied is separated from the photospheric radiation of $\eta$ 
Carinae, i.e. 
% 
% Eq 9 
\begin{equation} 
d\simeq\lambda_0 \frac{L}{fD}, 
\end{equation}
where $L = 10^{22}$ cm is the distance to $\eta$ Car, D$\sim 10^{15}$ cm is the diameter of 
the Weigelt blob, and f is the fraction of the blob region, in which the laser 
effect takes place to produce the radiation received by the ground-based 
telescopes. For estimation purposes, one can set f = 1, and the necessary 
distance between the telescopes will then be $d\sim 10^3$ cm. By choosing $d\sim 
10^4$ cm, one can, in principle, analyze laser radiation coming from blob regions 
as small as one-tenth of the size of the blob.

At the high-speed photomixer, the wavefront of the radiation being received 
should be matched with diffraction-limited accuracy to that of the local laser 
oscillator radiation over the entire area $S$ of the photodetector in accordance 
with the antenna theorem for photomixing (Letokhov, 1965; Siegman, 1966). In 
that case, $S\Omega\sim \lambda_0^2$, where $\Omega$ is the field of view or 
aperture. The heterodyne field of view will then be $\Theta\sim 
1.2\lambda_0/a_0$, where $a_0$ is the diameter of the primary telescope mirror. 
Modern avalanche photodiodes (Prochazka et al, 2004) with quantum efficiency $\eta\sim 1$ are most 
suitable to use as a high-speed photomixer. The photomixer signals of 
intermediate frequency $f = \nu_{\mathrm {het}} - \nu$, where $\nu_{\mathrm 
{het}}$ is the laser heterodyne frequency and $\nu$ is the frequency of the desired 
spectral line frequency of the astrophysical laser radiation being received, 
should be processed with a correlator. The measured correlation functions of the 
signal intensity fluctuations as a function of frequency (Fig. 6a) should 
provide information about the Fourier transformation of the spectral profile of 
the radiation being received. The spectral resolution $R$ will be determined by 
the reception bandwidth $B$ (Hz) of the photomixer radio signals:
%
% Eq 10
\begin{equation}
R= \frac{c}{\lambda_0B}.
\end{equation}
At $B\sim 10^6$ Hz the spectral resolution $R\sim 3\cdot10^8$, which is 
sufficient to measure the emission spectrum of an astrophysical laser with a 
spectral width tens of times narrower than the Doppler width. The dependence of 
the correlation signal on the distance $d$ between the telescopes should give 
information on the angular size $\varphi$ of the blob region wherein the astrophysical 
laser of interest is active at the wavelength $\lambda_0$ under study (Fig. 6b). However, 
the key problem for the experiment proposed is the signal-to-noise ratio S/N 
that poses the requirement for the primary mirror diameter $a_0$ of the telescopes, 
the reception bandwidth $B$, and the observation time $\tau$.

\begin{figure}
\begin{center}
    \rotatebox{270}{\resizebox{5cm}{!}{\includegraphics{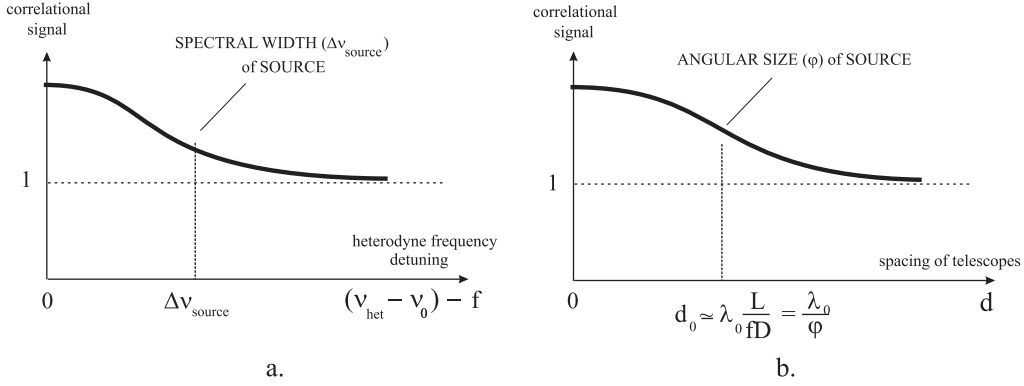}}}
\end{center}
\hfill  
     \caption{Expected dependence of the correlation signal intensity 
as a function of a) the heterodyne frequency detuning and b) the spacing of 
telescopes $d$. }
    \label{Fig. 6}
\end{figure}

\section{S/N Ratio for Optical Heterodyne APL Experiment}

With heterodyne detection, one can control the intensity of the local laser 
oscillator and thus raise the signal level above the noise level of the 
electronic circuits used. As a result, one can reach the quantum noise limit, at 
which the signal-to-noise level is given by (Abbas et al., 1976; Rothemal et 
al., 1983)
%
% Eq 11
\begin{equation}
\frac{S}{N}=\frac{\eta P}{{\mathrm h}\nu B},
\end{equation}
where $P$ is the power of the source, $B=\Delta\nu$ is the bandwidth of a single 
frequency resolution element of the mixing signal, and $\eta$ is the quantum 
efficiency of the photodetector-mixer. If the signal is integrated over time $\tau$
, the $S/N$ ratio is enhanced by a factor of $(B\tau)^{1/2}$, and becomes
%
% Eq 12
\begin{equation}
\frac{S}{N}=\frac{\eta P}{{\mathrm h}\nu B}\left (\frac{\tau}{B} \right )^{1/2}.
\end{equation}
The $S/N$ ratio drops with the increasing  
bandwidth $B=\Delta\nu$. To measure the profile of a spectral line having a 
width of tens and hundreds of MHz, i.e. much wider than $B$, it is not necessary 
to scan the entire profile with a spectral resolution of $B$. It is natural to 
suppose that the spectral profile of the APL radiation is bell-shaped. 
Therefore, it is sufficient to measure the signals within very narrow bands $B$, a 
few Hz wide, at several points of the profile and integrate over a sufficiently long 
time at each sampling point. In that case, one can achieve a 
sufficient increase in the $S/N$ ratio without any loss of essential information.

To estimate the $S/N$ ratio in the case of astrophysical laser lines from the 
Weigelt blobs in, one can use the HST/STIS data for the 
integrated intensity of one of these lines (Gull et al., 2001). For example, the 
intensity of the 9997 \AA\ line from the Weigelt blobs BD amounts to $2\cdot10^{–12}$ 
erg/cm$^2\cdot$s$\cdot$\AA. With the primary telescope mirror area $A\simeq a_0^2$
 of the order of 1 m$^2$, one can expect a power of $2\cdot10^{–8}$ erg/s within the 
limits of the Doppler profile $\Delta\nu_{\mathrm D}<<$ 1 \AA. From this we get 
a lower limit estimate of the $S/N$ ratio:
%
% Eq 13
\begin{equation}
\frac{S}{N}\geq 1.2\cdot 10^4\eta \left (\tau/B \right )^{1/2}.
\end{equation}
For $\eta \sim 1, \tau\sim 10^2$ s, and $B$= 1 kHz, we get $S/N\sim 4\cdot10^3$.

The estimate based on Eq. 13 should be considered very approximate, as it can be either 
greater or smaller by at least an order of magnitude for the following reasons. 
On one hand, the size of the laser volume in the blob can be smaller than 
that of the whole blob whose emission is received by the HST/STIS. Therefore, 
the integrated intensity of the 9997 \AA\ line can be lower than the value used for 
the estimation. On the other hand, the spectral width of the APL radiation 
can be an order of magnitude smaller than the Doppler width (Sect. 2). Furthermore 
, the divergence of the APL radiation in the case of an 
elongated volume for the amplifying medium can be much smaller than 4$\pi$. 
In any case, the estimates above point to the feasibility of the proposed experiment 
with the power of the radiation received varying over wide limits because of the 
existing variable parameter $(\tau/B)^{1/2}$. Let us underline that the 
proposed experiment can already be performed using existing pairs of ground-based 
telescopes in the Southern hemisphere.

\begin{ack}
V.L. acknowledges financial support through grants (S.J.) from the Royal Swedish Academy 
of Sciences and the Wenner-Gren Foundations. V.L is grateful to Lund Observatory for hospitality
and to the Russian Foundation for Basic Research for support through grant No 03-02-16377. 
The research project is also supported by a grant (S.J.) from the Swedish National Space Board. 
One of the authors (V. L.) is thankful to ESO (Dr. D. Alloin) for hospitality in Santiago, 
as part of the paper was written during his visit there. 
\end{ack}

\end{document}